\documentclass[12pt]{article}
\usepackage{a4}
\usepackage{amsfonts,amssymb,amsmath,amsthm}
\usepackage{graphicx}
\usepackage{slashed}
\usepackage{epsfig}
\usepackage{cite}
\usepackage{hyperref}
\def\({\left(}
\def\){\right)}
\def\[{\left[}
\def\]{\right]}

\def\om{\omega}
\def\be{\begin{equation}}
\def\e{\end{equation}}
\def\ee{\end{equation}}

\def\sP{\slashed\partial}
\def\s#1{\slashed{#1}}
\def\t{\tilde}

\def\Im{{\rm Im}}
\def\Re{{\rm Re}}

\begin{document}

\title{Parity-odd effects and polarization rotation in graphene}
\author{I.~V.~Fialkovsky${}^{*\ddag}$ and D.~V.~Vassilevich${}^{\dag,\ddag}$\\
\small ${}^*$Insituto de F\'isica, Universidade de S\~ao Paulo, SP, Brazil\\
\small ${}^\dag$CMCC, Universidade Federal do ABC, Santo Andr\'e, SP, Brazil\\
\small ${}^{\ddag}$Department of Theoretical Physics,
 St.Petersburg University, Russia}

\maketitle
\begin{abstract}
We show that the presence of parity-odd terms in the conductivity
(or, in other words, in the polarization tensor of Dirac
quasiparticles in graphene) leads to rotation of polarization of the
electromagnetic waves passing through suspended samples of
graphene. Parity-odd Chern-Simons type
contributions appear in external magnetic field, giving rise
to a quantum Faraday effect (though  other sources of parity-odd
effects may also be discussed). The estimated order of the effect is well above
the sensitivity limits of modern optical instruments.
\end{abstract}

 Since the experimental discovery of graphene \cite{expdis}, its
unusual properties are attracting growing attention of the
researchers, see reviews \cite{graprev}.
Quantum Field Theory proved to be a very successful approach to
description of dynamics of elementary excitations
in graphene which are supposed to
be $2+1$ dimensional fermions \cite{early}.
It is very well known,
that in the massless case such fermions exhibit the so-called parity
anomaly\footnote{For the first time the induced Chern-Simos term
was calculated by Redlich \cite{Redlich}. Later on, many issues
regarding the parity anomaly, especially at finite temperature,
were clarified in \cite{par}. For an overview, see \cite{Dunne}.},
which leads in particular to the presence of the Chern-Simons term
in the effective action for external electromagnetic field.
Other
parity-violating terms which can arise in the massive case $m\ne 0$,
enter the effective action multiplied by odd powers of $m$.
In graphene, the fermions are
doubled, and the corresponding gamma-matrices are taken in two
inequivalent representations (differing by an overall sign).
It is
possible therefore to adjust parameters the of the model
(couplings of the fermions to external field, etc.), and also the rules of analytic
continuation of quantum determinants in such a way that all
parity-odd terms in the complete effective action are precisely canceled.
Such cancellation, even if it happens, cannot be universal. Recent
works \cite{Arg} on Quantum Hall effect in graphene show that the
Chern-Simons term must appear at least in constant magnetic field.
It corresponds to the
non-diagonal (parity-odd) part of dynamical conductivity of
graphene in external magnetic field found in \cite{Gusynin06,Falkovsky08}.
One may even suppose that
such effects also appear under more general conditions. It is natural to
look for their physical manifestations.

In the present Letter we argue that the parity odd (quantum) effects
should lead to the polarization rotation of the electromagnetic
waves passing through suspended samples of mono- and few-layer
graphene films. Below, we
shall relate the polarization rotation angle to the real part of
the parity-odd one-loop two-photon amplitudes. We shall also
estimate the order of the effect and discuss physical conditions
under which this effect may be measurable.
It will definitely happen in an external magnetic field thus
giving rise to quantum Faraday effect, but the polarization
rotation without a magnetic field cannot be excluded as well. We
stress that both negative and positive result of an experiment
measuring such rotation will deliver valuable information on the
structure of graphene. To get an idea how large or small the effect
could be we shall consider the simplest model of a single massive
fermion.
We emphasize that the appearing P-odd contributions to
polarization operator, or, equivalently, additional terms in the
non-diagonal part of dynamical conductivity, do not
influence any predictions for quantum Hall effect in graphene since they
disappear in the dc limit ($\omega\to0\Leftrightarrow
m\to\infty$). On the other hand, parity-odd conductivity of any
nature unavoidably leads to polarization rotation effects, as is
evident from the calculations given below.

The mechanism of creation and physical manifestation of the Chern-Simons
like contributions to the effective action is very similar to those
proposed for 4-dimensional QED \cite{4D}, although in four dimensions
the effects appear at much larger (astrophysical) scales.

Optical polarization effects in graphite and
in arrays of carbon nanotubes
was subject to extensive study, see e.g. \cite{optpol}.
They were attributed to the anisotropy of the two-dimensional
Brillouin zone of graphite or, equivalently,
anisotropy of graphene crystal lattice. Similar ideas were used to predict orientation dependence of
absorption of high-frequency light in graphene \cite{Zhang08}.
The physics behind these effects is completely different from what we shall
consider below.

Let us formulate a model, or rather a class of models, we are going
to consider. We assume that the propagation of fermions is described
by a $2+1$ dimensional action
\begin{equation}
S_\psi = \int d^3x \, \bar\psi \(i\slashed\partial-e \s{A}+\dots \)\psi
\,, \label{Spsi}
\end{equation}
where
\be
    \sP\equiv\tilde\gamma^l \partial_l,\quad l=0,1,2
    \quad \tilde\gamma^0\equiv\gamma^0,\quad
\tilde\gamma^{1,2}\equiv v_F\gamma^{1,2},
        \quad \gamma_0^2=-(\gamma^1)^2=-(\gamma^2)^2=1\,.\label{gam}
\ee
The Dirac matrices
$\gamma^l$ may be taken in a reducible representation.
$v_F$ is the Fermi velocity.
In our units, $\hbar=c=1$, $v_F\simeq (300)^{-1}$.
The dots in (\ref{Spsi}) denote
other parameters, like the mass or chemical potential, see \cite{GSC}
for a rather complete list of allowed interactions and explanations of their
physical meaning. For us it is only
important that the interaction with the electromagnetic field $A$ is gauge
invariant.
The electromagnetic part of the
action we choose as $S_{{\it EM}}=-1/4\int d^4x F_{\mu\nu}^2$,
$\mu,\nu=0,1,2,3$.
In these units $e^2=4\pi\alpha\simeq 4\pi/137$.

After integrating out the fermions, one arrives
at the following effective action for the electromagnetic field in the
quadratic approximation
\be
    S_{\rm eff}=\frac 12 \int d^3xd^3y\, A_j(x) \Pi^{jl}(y-x) A_l(y) \,, \label{Seff}
\ee
where, after the Fourier transform,
\begin{equation}
    \Pi^{mn}
        = \frac \alpha{v_F^2} \, \eta^{m}_{j}\[
            \Psi(\tilde p) \(g^{jl}-\frac{\t p^j\t p^l}{\t p^2}\)
            + i \phi(\tilde p) \epsilon^{jkl}\t p_k
            \]\eta_l^n \label{Pmn}
\end{equation}
with $\epsilon^{012}=1$, $\eta_j^n={\rm diag}(1,v_F,v_F)$, $\t
p^m\equiv \eta_n^m p^n$. The functions  $\Psi$ and $\phi$ are
model-dependent and can take complex values. Some more comments
regarding the formula (\ref{Pmn}) are in order. From (\ref{gam})
it is clear that
due to the presence of effective 2-dimensional speed of light $v_F$,
quantum corrections shall depend on the rescaled
momentum $\tilde p$ rather than on $p$. This is reflected in the
construction in square brackets of (\ref{Pmn}) which reproduces
the standard tensor structure of the polarization tensor, but with
$p$ substituted by $\tilde p$. The multiplier $v_F^{-2}$ appears
due to the relation $d^3q=v_F^{-2}d^3\tilde q$ for the integration
measure for the loop momentum. The overall rescalings $\eta^m_j$
of the polarization operator appear since the electromagnetic
potential is also multiplied with rescaled gamma-matrices
(\ref{gam}). It worth mentioning, that the polarization tensor is transversal with respect to
the unrescaled momenta, $p_n\Pi^{nm}(p)=0$,
as expected from the gauge invariance of the model.

Precise form of the functions $\Psi(p)$ and $\phi(p)$ has been calculated
in various models. The function $\Psi$ was for the first time calculated in the
framework of 3d QED in \cite{Appelquist86}. In connection to
graphene systems it was presented in \cite{Gusynin02} for a
semi-relativistic reduced QED model containing three parameters
(mass gap, chemical potential and temperature). Later it was
generalized to include also the scattering rate and external
magnetic field in \cite{Gusynin06}, and recently rederived
in \cite{Pyatkovskiy09}. Predictions for the absorption of
light in the visible part of the spectra
based on calculations of the even part of the polarization function
(or, equivalently, of the dynamical conductivity \cite{Falkovsky08})
are in a good agreement with the experimental data \cite{exPeven}.
The parity-odd part of the polarization tensor for planar fermions
was calculated in \cite{Redlich,par}. Chern-Simons modifications
of the conductor boundary conditions were considered in \cite{EVB}.
The calculations of the Casimir force due to the Chern-Simons
action on a surface (without a parity-even part) were done
in \cite{Pi}.

Let us embed the surface occupied by the graphene sample
in the $3+1$ dimensional
Minkowski space by identifying it with the $x^3\equiv z=0$ plane.
The propagation of electromagnetic waves is described by
the Maxwell equations with a delta-function interaction
corresponding to (\ref{Seff})
\be
    \partial_\mu F^{\mu \nu}+\delta(z)\Pi^{\nu \rho} A_\rho=0\,,
    \label{maxw}
\ee
where we extended the polarization operator to a $4\times 4$ matrix
by setting $\Pi^{3\mu}=\Pi^{\mu3}=0$, $\mu,\nu=0,1,2,3$. Due to the
delta-function interaction  in (\ref{maxw}) the following matching condition
must be imposed on the field $A_\mu$
\be
\begin{array}{l}
  A_\mu\big|_{z=+0}=A_\mu\big|_{z=-0} \\
  (\partial_z A_\mu)_{z=+0}-(\partial_z A_\mu)_{z=-0}
    =\Pi_\mu^{\ \ \nu} A_\nu\big|_{z=0} \label{mc}
\end{array}
\ee

Let us consider a solution in the form of a plane wave
propagating along the $z$-axis
from $z=-\infty$ with the initial polarization parallel to $x^1\equiv x$,
which is being reflected by and transmitted through
the graphene sample.
\be
    A=e^{-i\om t}\left\{\begin{array}{ll}
        \mathrm{\bf e}_x e^{ik_3 z}
            + \(r_{xx}\mathrm{\bf e}_x
          +r_{xy}\mathrm{\bf e}_y\)e^{-ik_3 z}, & z<0 \\
        \(t_{xx}\mathrm{\bf e}_x +t_{xy}\mathrm{\bf e}_y\)e^{ik_3 z}, & z>0 \\
                \end{array}\right.\label{az}
\ee
where $\mathrm{\bf e}_{x,y}$ are unit vectors in the direction $x^{1,2}$.
The on-shell condition (free Maxwell equations
outside $z=0$) implies $k_3=\omega$.
For such waves the matching conditions (\ref{mc}) simplify,
\begin{eqnarray}
&& A_a\big|_{z=+0}=A_a\big|_{z=-0} \nonumber\\
&&  (\partial_z A_a)_{z=+0}-(\partial_z A_a)_{z=-0}
=\alpha \left[ \Psi (k) \delta_a^b + i \omega
\phi (k)\epsilon_{a}^{\ \ b} \right] A_b \,,\label{mc2}
\end{eqnarray}
where $a,b=1,2$, $\epsilon_1^{\ \ 2}=-\epsilon_2^{\ \ 1}=1$.
The transmission and reflection coefficients can be found
relatively easy
\begin{eqnarray}
&&t_{xx}=\frac{-2 \om (i\alpha \Psi+2\om)}{\alpha^2\Psi^2-4i\alpha \om\Psi
-(4+\alpha^2\phi^2)\om^2}\,,\nonumber\\
&&t_{xy}=\frac{2 \phi \alpha \om^2}{\alpha^2\Psi^2-4i\alpha \om\Psi
-(4+\alpha^2\phi^2)\om^2}\,,\nonumber\\
&& r_{xx}=t_{xx}-1,\qquad
    r_{xy}=t_{xy}\,.\label{ttrr}
\end{eqnarray}
The amplitude of the transmitted wave reads
\be
     \mathcal{A}=\sqrt{|t_{xx}|^2+|t_{xy}|^2}
        \simeq 1-\left| \frac{\Im\Psi}{2 \om}\right|{\alpha}
            +O(\alpha^2)\,,\label{Dt}
\ee
meaning that there is an absorption in the model
at the linear order in $\alpha$,
which is in agreement with experiment \cite{exPeven}. To all orders of $\alpha$
the flux conservation condition
$ |r_{xx}|^2+|r_{xt}|^2+|t_{xx}|^2+|t_{xy}|^2=1$ holds if both $\Psi$
and $\phi$ are real.

\begin{figure}\label{fig}
\includegraphics[width=4in]{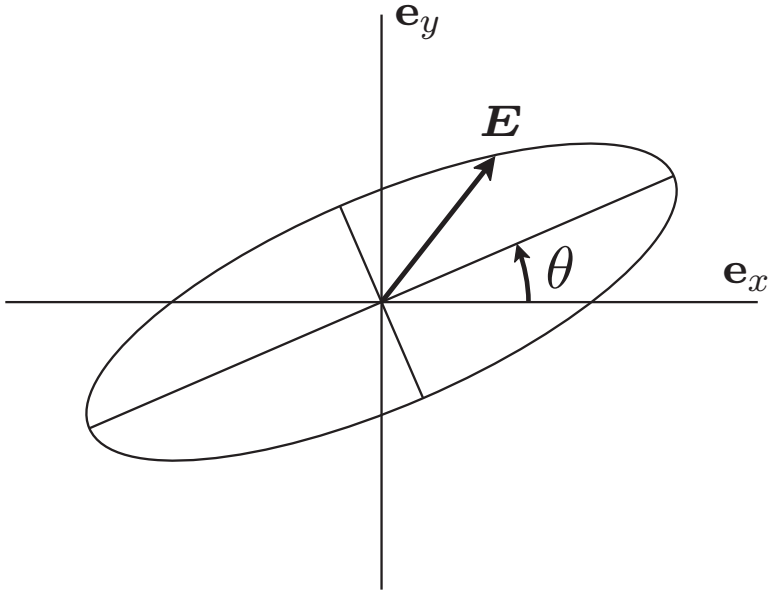}
\caption{Polarization of the transmitted wave}
\end{figure}

Let us study the polarization rotation. In generic case the transmitted
wave has an elliptic polarization, see Fig.~1. The ratio of the semiaxises
of the ellipse is given by
\begin{eqnarray}
&& R=\frac{|s_{+}|-|s_{-}|}{|s_{+}|+|s_{-}|}
        \simeq -\frac{\alpha \Im \phi}2+O(\alpha^2)
\label{rat}\\
&&
    s_{+}=\frac{t_{xx}-i t_{xy}}{2},\qquad
        s_{-}=\frac{t_{xx}+i t_{xy}}{2}
\nonumber
\end{eqnarray}
while the angle with ${\bf e}_x$ is
\be
    \theta
        =\frac12{\rm arg}\, \frac{s_+}{s_-}
=-\frac 12 {\mathrm{arg}}\, \frac {i\alpha\Psi + 2\omega +i\phi\alpha\omega}
{i\alpha\Psi + 2\omega -i\phi\alpha\omega}=
         -\frac{\alpha \Re\phi }2+O(\alpha^2)\,.
\label{theta}
\ee

Let us further note that
\begin{equation}
    t_{xy}=\frac{ -\alpha \om \phi}{i\alpha\Psi+2 \om}t_{xx}\,.
\label{tt}
\end{equation}
The transmitted wave is linearly polarized if the ratio $t_{xx}/t_{xy}$
is real. This happens exactly if $\Re\Psi=\Im\phi=0$, but in the leading order of
$\alpha$ it is enough to require $\Im\phi=0$. The angle between the electric
field of the transmitted wave and $\mathrm{\bf e}_x$ is still given by
(\ref{theta}).

To estimate the order of magnitude of the polarization rotation effects
let us consider
the simple case of a single massive fermion in $2+1$ dimensions.
The functions $\Psi$ and $\phi$ in (\ref{Pmn}) can be
calculated explicitly
\begin{eqnarray}
&&\Psi(p)=
        \frac{2 m \t p -(\t p^2+4m^2){\rm arctanh }({\t p}/{2m})}{2\t p},\qquad
    \label{Psi}\\
&&
\phi(p)
    =\frac{2m\, {\rm arctanh}(\t p/2 m )}{\t p}-1\label{phi}
\end{eqnarray}
here $\t p\equiv+\sqrt{\t p_j \t p^j}$, and we assume $m\geqslant0$. Both (\ref{Psi})
and (\ref{phi}) are consistent with earlier (and more general) computations,
see \cite{Redlich,par,Dunne, Appelquist86,Gusynin02,Gusynin06,Pyatkovskiy09}.
We remind that the term $-1$ in (\ref{phi}) is a result of
the Pauli-Villars subtraction which is necessary to restore full gauge
invariance of the effective action \cite{Redlich}.

Besides more elaborate couplings and larger number of parameters in realistic
models of graphene the number of fermionic components is larger. Therefore,
actual values of the functions $\Psi$ and $\phi$ may change significantly.
We believe, that the formulae are enough to make an order of magnitude
estimate of the expected effects.

For normally incident waves $\tilde p=\omega$. Both $\Psi$ and $\phi$ are real
for $\omega< 2m$ and complex for $\omega >2m$ yielding to absorption which
starts at $\omega =2m$. The absorption and polarization rotation effects
are small (so that one can use an $\alpha$-expansion) everywhere except
for a narrow region around $\omega = 2m$. At $\omega = 2m$ the angle
$\theta$ has a peak with a maximum value of $\pi/4$. It is likely however
that this large effect will be smeared out in a more realistic model.
At low frequencies, $\omega\ll2m$, the function $\phi$ vanishes and the
polarization rotation effects become negligible.

At large frequencies,
$\omega\gg2m$, $\Psi\simeq i\pi\tilde p/2$ and $\phi\simeq -1$.
Taking into account the actual number of fermion species used for
correct description of graphene quasi-particles, see
\cite{Gusynin06}, one finds that (\ref{Dt}) in high frequency regime
is in perfect agreement with experimental results
\cite{exPeven}. On the other hand, the rotation angle in this limit is
$\theta\simeq\alpha/2$ which is well above the sensitivity limit
of modern optical devices. Phenomenologically acceptable values of
the mass parameter (energy gap, chemical potential, etc.)
for many realistic systems is at maximum of order of $1 eV$.
Therefore, if a parity-odd part of the effective action exists,
the polarization rotation effects must be well detectable already in
the visible part of the spectrum.

Under which conditions could one expect the above described
effects to take place?
The first candidate to think of is a classically P-even system
in external magnetic field. In this case, the presence of
the Chern-Simons term and, consequently, of the polarization
rotation seems to be unavoidable. Indeed, the effective action for
fermions in graphene in the presence of a constant magnetic field
does contain a Chern-Simons term for this field \cite{Arg}. On the other hand,
direct calculations of the polarization operator in external magnetic
field (made in \cite{Gusynin06} for the case of vanishing spatial part of external momenta)
shows explicitly the Chern-Simon contribution
%(compare (\ref{Pmn}) with (A4) of \cite{Gusynin06}).
(compare RHS of (\ref{mc2}) effectively defining
the conductivity in our model with (A4), (A8) of \cite{Gusynin06}).
The polarization rotation due to parity-odd quantum effects in
external magnetic field is a kind of \emph{quantum Faraday
effect}.

An even more exciting (though a bit more speculative) possibility
is that parity-violating terms might appear without a constant external
magnetic field. Such a possibility was studied already in the
pioneering paper by Haldane \cite{Haldane:1988zza}.
Note, that even a relatively tiny non-cancellation of parity-odd
parts of the effective action between different fermion species,
as small as 10\% of (\ref{phi}),
should already be detectable with up-today experimental techniques.
Besides, there are 2-dimensional systems without reflection symmetry, like the
few-layer graphene films \cite{FLG06},
where there is no particularly good reason to
expect this cancellation at all.

To summarize, there are various arguments in favor and against
existence of parity-odd quantum effects in graphene and other
2-dimensional systems in condensed matter physics leading to
Chern-Simons type contributions to the effective action. This
situation may be resolved by simple experiments with the polarized
light transmitted through suspended films. Any result of such
experiments, positive or negative, will tell us a lot about the
electron structure of graphene and similar systems.

\bigskip

{\bf Acknowledgements}.
We are grateful to E.~M.~Santangelo for
correspondence. This work was
supported in part by FAPESP and CNPq.
I.F. also acknowledges financial support of the Russian Foundation
of Basic Research (grant RFRB $07$--$01$--$00692$).

\end{document}